\documentclass{article}
\usepackage{amsmath,mathtools}
\usepackage{amsfonts}
\usepackage{geometry}
\usepackage{graphicx}
\usepackage{bbold}
\usepackage{accents}
\usepackage{xcolor,soul}
\usepackage{multicol}
\usepackage{mdframed}
\usepackage{mathrsfs}
\usepackage{breqn}
\usepackage[colorlinks=true,urlcolor=blue,linkcolor=blue,citecolor=red]{hyperref}
\setulcolor{red}
\setul{1pt}{0.1em}
\def\beq{\begin{eqnarray}}  
\def\eeq{\end{eqnarray}} 

\def\sr{{\text{\tiny)}}}
\def\sl{{\text{\tiny(}}}
\def\ar{{\text{\tiny]}}}
\def\al{{\text{\tiny[}}}

\def\onehalf{{\textstyle{{\frac{1}{2}}}}}
\def\n{{\text{\tiny 0}}}
\begin{document}

\begin{center}
{\Large \bf On black holes in teleparallel torsion theories of gravity}
\vskip 0.7cm
{A. A. Coley\footnote{alan.coley@dal.ca}, N. T. Layden\footnote{nicholas.layden@dal.ca} and D. F. L\'opez\footnote{diego.lopez@dal.ca} }
\vskip 0.2cm
{\it Department of Mathematics and Statistics\\ 
Dalhousie University\\
Halifax, Canada}
\vskip 0.8cm
\end{center}

\begin{abstract}
We first present an overview of the Schwarzschild vacuum spacetime within general relativity, with particular emphasis on the role of scalar polynomial invariants and the  null frame approach (and the related Cartan invariants), that justifies the conventional interpretation of the Schwarzschild geometry as a black hole spacetime admitting a horizon (at $r=2M$ in Schwarzschild coordinates) shielding a singular point at the origin. We then consider static spherical symmetric vacuum teleparallel spacetimes in which the torsion characterizes the geometry, and the scalar invariants of interest are those constructed from the torsion and its (covariant) derivatives. We investigate the Schwarzschild-like spacetime in the teleparallel equivalent of general relativity and find that the torsion scalar invariants (and, in particular, the scalar $T$) diverge at the putative ``Schwarzschild'' horizon. In this sense the resulting spacetime is {\em not} a black hole spacetime. We then briefly consider the Kerr-like solution in the teleparallel equivalent of general relativity and obtain a similar result. Finally, we investigate static spherically symmetric vacuum spacetimes within the more general $F(T)$ teleparallel gravity and show that if a such a geometry admits a horizon, then the torsion scalar $T$ necessarily diverges there; consequently in this sense such a geometry also does {\em not} represent a black hole.
\end{abstract}
\vskip 0.5cm

\newpage	
%
%
\section{Introduction}

In Einstein’s theory of general relativity (GR), the spherically symmetric vacuum Schwarzschild solution, which by Birkhoff’s theorem is the only static spherically symmetric and asymptotically flat solution to the Einstein field equations (FE), gives rise to black hole solutions in which there is a horizon.  In GR the geometry is characterized by the curvature of the Levi-Civita connection which is derived from the metric. 

In teleparallel gravity (TG), the geometrical quantity of interest is the torsion, which is computed from the frame (or co-frame) and a zero curvature, metric compatible, spin connection. Understanding this alternative framework involves studying static spherically and axially symmetric geometries, particularly their vacuum solutions, which are crucial for assessing the viability of gravitational theories. These describe non-rotating and rotating black holes, known as the Schwarzschild and Kerr solutions, in the context of GR. However, these solutions present intriguing phenomena within the teleparallel context, challenging traditional ideas about black holes and gravitational singularities; in contrast to GR, where singularities typically occur only at the origin of the radial coordinate, TG exhibits divergent behaviour near both the horizon and origin.

The teleparallel equivalent of GR (TEGR) is a subclass of TG torsion theories based on the scalar, $T$, called the torsion scalar, that appears in the action. This theory is locally dynamically equivalent to GR, meaning their FE and solutions, including the Schwarzschild and Kerr solutions, are identical \cite{aldrovandi1732013}. However, the theories differ conceptually and geometrically, affecting their possible descriptions as black holes. While black holes are well-defined in GR, TEGR lacks a clear description of these spacetime regions. In this paper we wish to present a comprehensive study of possible black holes in TEGR using a fully covariant approach.

New general relativity (NGR) is a straightforward modification of GR, defined by including additional torsion scalars in the action through three free parameters. It contains TEGR as a special case where $T$ is constructed from a particular linear combination of these scalar invariants of the torsion tensor \cite{aldrovandi1732013}. In its original Lorentz-restricted formulation \cite{hayashi191981}, NGR provides a spherically symmetric vacuum exact solution for the theory's three-parameter family. However, this solution has been shown to describe a geometry with a singular horizon \cite{wanxiu1988,kawai831990,chengmin1995}. Using a fully invariant approach, it was further proven that more general spherically symmetric solutions of NGR produce torsion invariants that still display two singularities \cite{obukhov672003}. Modern approaches have confirmed these findings and discovered new types of solutions, though the issue of the singular horizon remains unresolved \cite{asukula2311,golovnev412024}. The development of axially symmetric vacuum solutions in NGR, as a generalization of the spherically symmetric case, has followed a similar evolution. Using the Lorentz-restricted approach of NGR, \cite{fukui1981} presented a solution that depended on one parameter and included the Kerr solution. The singularities of the Kerr-like spacetime were analyzed in \cite{toma1991,hecht1992}. In the future we will conduct a comprehensive study of exact solutions to determine if NGR admits any black holes  \cite{diego}.

Another important generalization of TEGR is $F(T)$ TG, where $F$ is an arbitrary twice differentiable function of the torsion scalar $T$ appearing in the action \cite{bahamonde862023, cai792016, ferraro752007, krssak362019}. When the geometrical framework describing $F(T)$ TG is defined in a gauge invariant manner, the resulting FEs derived from such a theory are fully Lorentz covariant \cite{krssak362019, KS2016}, and there will exist a frame/spin connection pair in which the spin connection identically vanishes (in this ``proper'' frame) \cite{aldrovandi1732013, krssak362019}. Regarding the static spherically symmetric vacuum solutions, a general family of such solutions has been obtained in the weak approach to $F(T)$ in \cite{pfeifer2021}. In \cite{coley616439,coley842024}, by employing a novel method to determine general geometries with spherical symmetry, a range of new solutions was presented. Affine frame symmetries with a non-trivial linear isotropy group were studied in \cite{coley616439,hohmann2019modified}. In particular, in \cite{coley616439}  the frame/spin connection pair that describes the most general spherically symmetric teleparallel geometries was presented. Additionally, \cite{bahamonde2021general} examined spherically symmetric vacuum solutions for a generalization of NGR called $f(\mathscr{V},\mathscr{A},\mathscr{T})$ theory, which incorporates an arbitrary function $f$ of the irreducible parts of the torsion, $(\mathscr{V},\mathscr{A},\mathscr{T})$, in the Lagrangian theory definition. Despite all of these efforts, no one has provided a non-trivial exact black hole solution, except in cases with topologically flat horizons or complex-valued tetrads, which are problematic from a physical standpoint \cite{golovnev412024}. The $F(T)$ approach to axially symmetric vacuum solutions yields new solutions beyond the Kerr solution; however, these solutions still exhibit a singular horizon \cite{jarv2019flat,bahamonde2021exploring}.

Static spherically symmetric solutions in $F(T)$ theory were reviewed in \cite{bahamonde862023}. We note that the Birkhoff theorem does not generally hold in $F(T)$ theories. Finding exact vacuum static spherially symmetric $F(T)$ TG solutions is an open problem, and there are no known exact vacuum black hole solutions. One exact non-trivial vacuum solution in $F(T)$ TG is the non-black hole power law solution presented in \cite{golov1}. There are a number of special static spherically symmetric (anisotropic) non-vacuum solutions known \cite{bahamonde862023}, including both black hole and non-black hole solutions (cf.  \cite{awad1,baha6,baha4}).

\subsection{Schwarzschild geometry}

When Schwarzschild originally obtained his solution for static, spherically symmetric vacuum metrics to the Einstein FE, the status of the ``singularity" at $r = 2M$ was first not understood  \cite{Unruh}. 
Indeed, at first it was regarded as a ``firestorm"; however, it was known at an early time that the status of singularities in a metric was not clear since singularities could result from a singular choice of coordinates. When Lema\^{i}tre \cite{referenceC8} found a coordinate transformation that regularized the Schwarzschild metric across the horizon, it was explicitly understood that the ``singularity" in the metric was consequently an artifact introduced because of the coordinates that Schwarzschild had used. A number of  coordinate transformations are now known which are regular in the whole of the extended Schwarzschild spacetime. 

Indeed, Eddington \cite{referenceC4} discovered an explicit coordinate transformation which is regular at $r = 2M$, which was rediscovered by Finkelstein \cite{referenceC5} who clearly recognized that this implied that the  null form of the Schwarzschild singularity (in Eddington-Finkelstein coordinates) was purely a coordinate artifact.  
The resulting metric is regular, and comes in two forms which cover different regions of an extended spacetime (but covers the exterior $(r>2M)$ region).

There exist coordinate systems for which the only singularity occurs at $r=0$, and in which the metric is otherwise regular and complete. Perhaps the best known is the Szekeres-Kruskal metric \cite{referenceC10}.
Israel \cite{Israel} (see also \cite{Newman}), found another global coordinate system, which profits from the fact that  that the coordinate transformation can be expressed with the new coordinates being explicitly given in terms of the Schwarzschild coordinates (not implicitly as in the Kruskal form). One of the coordinates $U$ is the same as the null coordinate of the Kruskal metric, while $z = const$ is a timelike hypersurface unless $z = 0$ whence it is a hypersurfaces. The metric in these Israel-Newman-Pajerski (INP) global coordinates is given and utilized below.

\subsection{Black holes}

We shall assume here that black holes are geometries that admit a horizon that shields a spacetime singularity \footnote{In principle, we need to define precisely what is meant by a black hole in alternative geometries (e.g. in the presence of torsion). This is one of the points of the current work. Here we emphasize the existence of a horizon (and that a spacetime is regular outside of it). Although, we consider the existence of a singularity shielded by the horizon as part of the definition of a black hole, the results obtained are valid regardless of what happens in the interior region of the horizon (and, for example apply to so called "singularity-free black holes") \cite{R2}.}.
The surface $r=2M$ is a horizon in the Schwarschild manifold. It is a global event horizon, but usually it is useful to characterize it locally in terms of an apparent horizon (AH) \cite{refAH}. A {geometric horizon} (GH) can be invariantly defined by the vanishing of a particular set of  scalar polynomial curvature invariants  \cite{GH}, which are frame independent.
In the case of spherically symmetric black holes, a GH  is equivalent to a AH \cite{GH}. 

The definition of a GH can also be given in terms of Cartan invariants \cite{kramer}, which must be calculated in a certain prescribed invariant coframe within the Newman-Penrose (NP) formalism. A GH implies that $\rho = 0$, defining an expansion free surface (where $\rho$ (and $\mu$) is a NP spin coefficient). Indeed, the conditions for an AH can be stated in terms of the Cartan invariants \cite{GH}:
\begin{equation} \rho = 0,~ \mu >0, ~~ \Delta \rho > 0, \label{CK_AH_conditions}
\end{equation} 
which replaces the phenomenological equation often used for
locating the AH \cite{Hochberg:1998ha}:  $g^{ab}R_{,a}R_{,b} =0$, where $R$  is the areal radius. 

It is also of interest to study the geodesic structure of a manifold, especially in the vicinity of the horizon. In the teleparallel case, this is done entirely via the metric, as in the Riemannian case.
For stationary black hole solutions, the AH can be identified with the Killing horizon, and so it is of interest to investigate the Killing vectors of the metric and examine the null hypersurfaces generated by them.

In addition, scalar curvature invariants, such as the Ricci scalar, Kretschmann scalar and Weyl squared scalar (or equivalently Cartan scalars), can be used to examine the curvature structure of a putative black hole. Such an approach has recently been used in the study of possible black hole spacetimes in Brans-Dicke theory \cite{nick}.
Note that in a Riemannian geometry, if one scalar curvature invariant constructed algebraically from the Riemann tensor and its covariant derivatives
diverges, it is singular there. 
We shall study teleparallel geometries by investigating scalar curvature invariants constructed algebraically from the torsion tensor and its covariant derivatives.

\subsection{Overview}

In this paper, we shall first review the necessary mathematical preliminaries and the use of Cartan invariants. We briefly review the Schwarzschild solution in conventional coordinates, then present the analysis in global analytic coordinates and provide an overview of the geodesic structure. We then discuss TEGR and study the TEGR Schwarzschild-like solution using invariants. After that, we briefly describe the TEGR Kerr-like solution. We then consider vacuum static spherically symmetric spacetimes in $F(T)$ theory. Finally, we conclude with some remarks.

Let us briefly review our notation conventions. We will use \( e \) to denote the frame, \( g \) for the spacetime metric, and \(\eta\) to denote both the null and Minkowski metrics  in their respective contexts. To differentiate between the two geometrical setups presented here, we will use \( \gamma \) and \( \Gamma \) for the spin and Levi-Civita connections in GR associated with the covariant derivative denoted by \( ; \). The Riemann tensor, Ricci tensor, Einstein tensor, and Kretschmann scalar will be denoted by \(\mathcal{R} \), \(  \mathcal{G} \), and \( \mathcal{K}\), respectively. Conversely, we will use \( \omega \) and \( \Omega \) for the spin and teleparallel connections in TG associated with the covariant derivative denoted by \( | \). The torsion tensor and the torsion scalar in particular will be denoted by \( T \), and the contortion tensor by \( K \).

%
%
\section{Mathematical preliminaries}

\subsection{Newman-Penrose formalism}

We present a brief overview of the frame formalism and an application to the exterior and global Schwarzschild solutions. A very convenient way of approaching problems in GR is to recast the problems written in terms of coordinate bases into ones written in terms of a frame of vectors associated with the metric \cite{kramer}. For the analysis in GR, we employ the NP formalism, and later relate it to a similar approach in TEGR.
		
Given a NP null frame $\{ \ell,n,m,\bar{m}\}$ (a set of four \textit{null} vectors) and its corresponding dual coframe, we define the tetrad (sometimes referred to as the \textit{vierbeins}), the components of the null frame, and their inverse, as:
\begin{equation}
\begin{aligned}
e_{a}{}^{\mu}&=\{ \ell^{\mu},n^{\mu},m^{\mu},\bar{m}^{\mu}\}^T, \\
e^{a}{}_{\mu}&=\{ -n_{\mu}, -\ell_{\mu}, \bar{m}_{\mu}, m_{\mu}\}\, .
\end{aligned}
\end{equation}
The Latin indices denote indices on the null frame (``tetrad indices"), while Greek indices denote standard spacetime indices (``coordinate indices") in the coordinate frame. A frame or co-frame is considered dual (or orthogonal) if it satisfies the conditions \( e_{a}{}^{\mu} e^{b}_{\mu} = \delta_{a}^{b} \) and \( e_{a}{}^{\nu} e^{a}{}_{\mu} = \delta_{\mu}^{\nu} \). These orthogonality conditions can be easily verified using the null frame conditions:
	\begin{equation}
		\begin{aligned}
			-\ell_{\mu} n^{\mu}=m^{\mu} \bar{m}_{\mu}=1, \\
			\ell_{\mu} \ell^{\mu} = n_{\mu} n^{\mu} = m_{\mu} m^{\mu} = \bar{m}_{\mu} \bar{m}^{\mu} = 0,
		\end{aligned}
	\end{equation}
and all other inner products of the null vectors vanish. We have chosen the null frame vectors such that the metric of the spacetime is given in terms of the null vectors and vierbein as\footnote{Brackets surrounding indices denote the symmetrization of the indices: $\ell_{\sl\mu}n_{\nu\sr}=\frac12(\ell_{\mu}n_{\nu} + \ell_{\nu} n_{\mu}).$}:
\begin{equation}\label{gmetric}
	g_{\mu\nu} = -2 \ell_{\sl\mu}n_{\nu\sr} + 2m_{\sl\mu}\bar{m}_{\nu\sr} = 	\eta_{ab} e^{a}{}_{\mu} e^{b}{}_{\nu} \, ,
\end{equation}
where the null metric components $\eta_{ab}$ are given by:
\begin{equation}\label{eta}
	\eta_{ab}=\left(\begin{matrix}
		0 & -1 & 0 & 0\\
		-1 & 0 & 0 & 0\\
		0 & 0 & 0 & 1\\
		0 & 0 & 1 & 0
	\end{matrix}\right) \, .
\end{equation}
For tensors, with respect to the null frame we raise and lower indices using $\eta$ as defined in (\ref{eta}). And for objects in the usual coordinate frame, we raise and lower indices using the spacetime metric, $g$, given in (\ref{gmetric}).

\subsection{Connection coefficients}

Given a frame basis defined by our tetrad, we compute the connection coefficients by calculating the covariant derivatives of all basis vectors and forming the connection coefficients for the null frame. These coefficients are known as the Ricci rotation coefficients (or spin coefficients in the case of an NP tetrad):
\begin{equation}
	\gamma^{c}{}_{ab} = e^{c}{}_{\nu}e_{b}{}^{\mu}e_{a}{}^{\nu}_{~;\mu},
\end{equation}
which are manifestly antisymmetric in the first two indices, which easily verified by taking a covariant derivative of the duality conditions described previously. Using this definition for the Ricci rotation coefficients, we can expand the covariant derivative in the coordinate frame, and obtain a relationship between the rotation coefficients, $\gamma$, and the standard Christoffel symbols, $\Gamma$:
\begin{equation}
	\gamma^{c}{}_{ab}=e^{c}{}_{\nu}\,e_{b}{}^{\mu}(\partial_{\mu}e_{a}{}^{\nu}+		\Gamma^{\nu}{}_{\mu\rho}\,e_{a}{}^{\rho})\,.
\end{equation}
By inverting this expression to isolate the Christoffel symbols, we arrive at an expression that gives the connection coefficients of the coordinate frame, directly in terms of the spin coefficients of the null frame:
\begin{equation}\label{connectiontospin}
	\left(\gamma^{c}{}_{ab}\,e_{c}{}^{\nu}\,e^{b}{}_{\mu}-\partial_{\mu} 			e_{a}{}^{\nu}\right)e^{a}{}_{\rho}=\Gamma^{\nu}{}_{\mu\rho}.
\end{equation}
Another useful way to view this equation is to expand the left hand side to obtain a projection of the spin coefficients onto the coordinate frame plus an extra term involving partial derivatives of the frame vectors:
\begin{equation}
	\Gamma^{\nu}{}_{\mu\rho}=\left(\gamma^{c}{}_{ab}\,e_{c}{}^{\nu}\,e^{b}{}		_{\mu}\,e^{a}{}_{\rho}\right) + \left(\ell_{\rho}\,\partial_{\mu}n^{\nu} + 		n_{\rho}\,\partial_{\mu}\ell^{\nu} - m_{\rho}\,\partial_{\mu}\bar{m}^{\nu}-		\bar{m}_{\rho}\,\partial_{\mu}m^{\nu}\right).
\end{equation}
\subsection{Spin coefficients}

In the NP formalism, we define the so-called spin coefficients by linear combinations of the Ricci rotation coefficients of the null frame, in the following manner (we have used a metric with signature ($-+++$), in accordance with the definitions by Kramer et al. \cite{kramer}). Up to potential sign differences, the components coincide with the definitions by Chandrasekhar \cite{refChandrasekhar}. In this set of equations, we specifically use Latin indices to not have the notation clash with that of the spin coefficient labels, but each scalar is defined in terms of the coordinate frame:
\begin{subequations}
\setlength{\abovedisplayskip}{0pt}
\setlength{\belowdisplayskip}{0pt}
\setlength{\columnsep}{4em}
\begin{multicols}{2}
\vspace*{-2\baselineskip}
\begin{flalign}
\qquad
-\kappa &\equiv m^{a}\ell^{b}\ell_{a;b} \, ,
&&\\
 -\rho &\equiv  \bar{m}^{a}\ell^{b}m_{a;b}\, ,
&&\\
-\sigma &\equiv m^{a}m^{b}\ell_{a;b}  \, ,
&&\\
-\tau &\equiv  m^{a}\ell^{b}\ell_{a;b} \, ,
&&\\
\nu &\equiv \bar{m}^{a}n^{b}n_{a;b}\, ,
&&\\
\mu &\equiv \bar{m}^{a} m^{b}n_{a;b}\, ,
&&
\end{flalign}
\begin{flalign}
\qquad
\lambda &\equiv \bar{m}^{a}\bar{m}^{b}n_{a;b}\, ,
&&\\
\pi &\equiv \bar{m}^{a}\ell^{b}n_{a;b}\, ,
&&\\
-\epsilon &\equiv \onehalf(m^{a}\ell^{b}\bar{m}_{a;b}-n^{a}\ell^{b}\ell_{a;b})    \, ,
&&\\
-\gamma &\equiv \onehalf (\ell^{a}n^{b}n_{a;b}-m^{a}n^{b}\bar{m}_{a;b})    \, ,
&&\\ 
-\beta &\equiv \onehalf (n^{a}m^{b}\ell_{a;b}-\bar{m}^{a}m^{b}m_{a;b})     \, ,
&&\\
\alpha &\equiv \onehalf (\ell^{a}\bar{m}^{b}n_{a;b}-m^{a}\bar{m}^{b}\bar{m}_{a;b}) \, .
&&
\end{flalign}
\end{multicols}
\end{subequations}
\subsection{Lorentz transformations}

The real usefulness of a null tetrad, like in the NP formalism, is in the very convenient form that the Lorentz transformations take. There are only three classes of Lorentz transformations, and they apply to the frame vectors in a simple manner:
\begin{enumerate}
	\item Null rotations about $\ell$ or $n$, with complex functions 			$E(x^{\mu}),B(x^{\mu})\in \mathbb{C}$.
	\item Boost transformations on $\ell$ or $n$, with $\lambda_{b}(x^{\mu})	\in \mathbb{R}$.
	\item Spin transformations about $m$ and $\bar{m}$, with                	$\Theta(x^{\mu}) \in \mathbb{R}$.
\end{enumerate}

In total, we have six transformation parameters, two are the real valued functions $\lambda_b,\Theta$, and the other four are the real and imaginary parts of the complex functions $E,B$. Writing the transformation rules for any of the NP scalars or spin coefficients under any Lorentz transformation thus becomes a simple task. A completely general Lorentz transformation of an NP frame $\{\ell^{\mu},n^{\mu},m^{\mu},\bar{m}^{\mu}\}$ is given by:
\begin{subequations}
	\begin{align}
	{\ell}^{\prime \mu} &= \lambda_b \ell^{\mu}+e^{i\Theta}\bar{B}(E 			\ell^{\mu}+m^{\mu})+e^{-i\Theta}B(\bar{E}\ell^{\mu}+\bar{m}^{\mu})+			\frac{B\bar{B}}{\lambda_b}(\bar{E}m^{\mu}+n^{\mu}+E(\bar{E}					\ell^{\mu}+\bar{m}^{\mu}))\,, \\
	{n}^{\prime \mu}&=\frac{1}{\lambda_b}\left(n^{\mu}+\bar{E}m^{\mu}+ 			E(\bar{E}\ell^{\mu}+\bar{m}^{\mu})\right)\,, \\
	{m}^{\prime \mu}&=e^{i \Theta}(E\ell^{\mu}+m^{\mu})+\frac{B}{\lambda_b}		\left(n^{\mu}+\bar{E}m^{\mu}+E(\bar{E}\ell^{\mu}+m^{\mu})\right)\,, \\
	{\bar{m}}^{\prime \mu}&=e^{-i \Theta}(\bar{E}\ell^{\mu}+\bar{m}^{\mu})+ 	\frac{\bar{B}}{\lambda_b}\left(n^{\mu}+\bar{E}m^{\mu}+E(\bar{E}				\ell^{\mu}+\bar{m}^{\mu})\right).
	\end{align}
\end{subequations}
Swapping $E \leftrightarrow B$, $\ell \leftrightarrow n$ gives the equivalent transformation with the order of the null rotations changed.
	
%
%
\section{Exterior Schwarzschild coordinate systems}

The standard ``Schwarzschild" coordinate system, with coordinates denoted by $(t,r,\theta,\phi)$, is used in writing down the exterior Schwarzschild metric given by the line element:
\begin{equation}\label{schm}
	ds^2 = -\left( 1- \frac{2M}{r}\right) dt^2 + \left( 1 - \frac{2M}{r} 		\right)^{-1} dr^2 + r^2(d\theta^{2}+\sin^{2}\theta \,d\phi^{2}),
\end{equation}
with $M \in \mathbb{R}$ is the Schwarzschild mass. This metric is valid for $r>2M$. Using the frame formalism, we define a null coframe given by the four vectors (in components):
\begin{subequations}\label{frameGR}
	\begin{align}
	\ell_{\mu}&=\frac{1}{\sqrt{2}}\left\{\sqrt{1-\frac{2M}{r}},\frac{1}{\sqrt{1-\frac{2M}{r}}},0,0\right\},\\
	n_{\mu}&=\frac{1}{\sqrt{2}}\left\{\sqrt{1-\frac{2M}{r}},-\frac{1}{\sqrt{1-\frac{2M}{r}}},0,0\right\}, \\
	m_{\mu}&= \frac{1}{\sqrt{2}}\left\{ 0,0, r, i r\sin(\theta)\right\}, \\
	\bar{m}_{\mu}&=\frac{1}{\sqrt{2}}\left\{0,0,r,-ir\sin(\theta)\right\}.
	\end{align}
\end{subequations}
This frame defines the invariant null frame associated with the Schwarzschild solution, and using it we compute the following NP scalars, the (extended) Cartan scalars:
\begin{subequations}
	\begin{align}
		\rho &= \mu = -\frac{\sqrt{1-2M/r}}{\sqrt{2} r}, \\
		\epsilon &= \gamma = \frac{M}{2\sqrt{2}r^2 \sqrt{1-2M/r}}, \\
		\beta &= -\alpha = \frac{\cot(\theta)}{2\sqrt{2} r}.
	\end{align}
\end{subequations}
The NP curvature scalar is:
\begin{subequations}
	\begin{align}
		\Psi_2 = -\frac{M}{r^3}.
	\end{align}
\end{subequations}
The frame derivatives of $\Psi_2$ are also given simply via the Ricci and Bianchi identities:
\begin{equation}
	D\Psi_2 =-\Delta\Psi_2=3\Psi_2 \rho = -\frac{3}{\sqrt{2}}\frac{M}{r^4}\sqrt{1-\frac{2M}{r}}.
\end{equation}
All other NP quantities associated with the frame are zero. We also give the Kretschmann scalar:
\begin{subequations}\label{Ksch}
	\begin{align}
		\mathcal{K} &= \mathcal{R}^{\alpha\beta\mu\nu}\mathcal{R}_{\alpha\beta\mu\nu}=48\Psi_2^{~2} = \frac{48M^2}{r^6}\, .
	\end{align}
\end{subequations}
	
We note that all scalar invariants constructed from the Riemann tensor are finite at the horizon. The spin coefficients $\epsilon = \gamma$ above actually diverge there. Therefore, it is necessary to check whether scalar invariants constructed from the covariant derivative(s) of the Riemann tensor are well behaved at the horizon (they are) and the geodesic structure there (see below).

\subsection{Global Schwarzschild coordinate systems}
	
We can define a global coordinate system for the Schwarzschild solution with coordinates which are explicited related to the exterior coordinates (but are also analytic in the interior). We present the Schwarzschild metric in terms of these coordinates, first given by Israel \cite{Israel}, and then separately by Newman and Pajerski \cite{Newman}. We denote these coordinates the ``INP" coordinate system, and relate it to the standard Schwarzschild coordinates above. For a comprehensive discussion of the INP coordinates, and their qualities, see the paper by Bill Unruh on his website \cite{Unruh}. The line element in INP coordinates is given by:
\begin{equation}
	ds^2 = \left(\frac{z^2}{2Mr}\right)du^2 + 2dudz + r^2 (d\theta^2 + 			\sin^2(\theta) d\phi^2 )\,,
\end{equation}
where $r=2M+uz/4M$. The following relations transform between INP and SC coordinate systems:
\begin{subequations}
	\begin{align}
	&t = 2M+\frac{uz}{4M}-2M \ln \left( \frac{u}{2z}\right ),\\
	&r = 2M+\frac{uz}{4M}\,.
	\end{align}
\end{subequations}
We can derive the following invariant null frame for the Schwarzschild metric in this coordinate system (via the Cartan-Karlhede algorithm):
\begin{subequations}
	\begin{align}
	&\ell^{\mu}=\left\{0,2\sqrt{2}\sqrt{\frac{M^2z}{u(8M^2+uz)}},0,0 \right\},\\
	&n^{\mu}=\left\{-\frac{1}{2\sqrt{2}}\sqrt{\frac{u(8M^2+uz)}{M^2 z}},\frac{1}{2\sqrt{2}}\frac{z^2}{(8M^2+uz)}\sqrt{\frac{u(8M^2+uz)}{M^2 z}},0,0\right\}, \\
	&m^{\mu}=\left\{ 0,0,\frac{1}{\sqrt{2}(2M+uz/4M)},\frac{i}{\sqrt{2}\sin(\theta)(2M+uz/4M)}\right\}, \\
	&\bar{m}^{\mu}=\left\{ 0,0,\frac{1}{\sqrt{2}(2M+uz/4M)},\frac{-i}{\sqrt{2}\sin(\theta)(2M+uz/4M)}\right\}. 		
	\end{align}
\end{subequations}
In this invariant frame, the spin coefficients have the following expressions and relations:
\begin{subequations}
	\begin{align}
	&\mu = \rho = -\frac{2\sqrt{2}Mu}{8M^2+uz}\sqrt{\frac{z}{u(8M^2+uz)}},\\
	&\gamma=\epsilon= \frac{4\sqrt{2}M^3}{u(8M^2+uz)^2}\sqrt{\frac{u(8M^2+uz)}{z}}, \\
	&\beta=-\alpha=\frac{\sqrt{2}M\cot(\theta)}{8M^2+uz}.
	\end{align}
\end{subequations}
Note that it is a coincidence that this frame has the same spin coefficients and relations as the Schwarzschild coordinates frame, as they are the extended invariants, not the actual Cartan scalar invariants.
The only nonzero NP curvature scalar is given by $\Psi_2$ (this is the only Cartan scalar invariant; as in the Schwarzschild coordinates system, all other extended scalars can be written in terms of $\Psi_2$):
\begin{equation}
	\Psi_2 = -\frac{64M^4}{(8M^2+uz)^3} = -\frac{M}{(2M+uz/4M)^3}.
\end{equation}
The first derivative of the Weyl tensor contains only components which are proportional to $D\Psi_2$:
\begin{equation}
	D\Psi_2=-\Delta \Psi_2=3\rho \Psi_2 = \frac{384\sqrt{2}M^5u}{(8M^2+uz)^4}\sqrt{\frac{z}{u(8M^2+uz)}}.
\end{equation}
Again, we can also compute the scalar polynomial invariant:
\begin{subequations}
	\begin{align}
	\mathcal{K} &= \mathcal{R}^{\alpha\beta\mu\nu}\mathcal{R}_{\alpha\beta\mu\nu} = 48\Psi_2^{~2}=\frac{48M^2}{(2M+uz/4M)^6}\, ,
	\end{align}
\end{subequations}
where we notice, as should be the case, that the scalar $\mathcal{K}$ has no singularities apart from the one at the origin $r=0$. The benefit here is that now the coordinates cross the event horizon, we can examine the behaviour of the scalars, and the geodesic equations as null geodesics cross the event horizon.
\subsection{Global analytic geodesic equations for Schwarzschild}

We can write the geodesic equations in terms of a null frame spin connection by transforming the Christoffel symbols (the connection coefficients derived from the metric directly) into the Ricci rotation coefficients, which are the analogous object derived from a null frame for the metric. 

Let us first consider the geodesic equations in terms of the spin coefficients. Given a null tetrad, $e_{b}{}^{\mu}$, and the spin coefficients for the null frame, $\gamma^{c}{}_{ab}$, we write the geodesic equations below in the original form, then in the `null basis' form: 
\begin{equation} \label{geodesiceqn}
	\frac{d^2x^{\rho}}{d\lambda^2}+\Gamma^{\rho}{}_{\mu\nu}\frac{dx^{\mu}}{d\lambda}\frac{dx^{\nu}}{d\lambda}=0\,,
\end{equation}
\begin{equation}
	\frac{d^2x^{\rho}}{d\lambda^2}+\left(\gamma^{c}{}_{ab}\,e_{c}{}^{\rho}e^{b}{}_{\mu}-\partial_{\mu}e_{a}{}^{\rho}\right)e^{a}{}_{\nu}\frac{dx^{\mu}}{d\lambda}\frac{dx^{\nu}}{d\lambda}=0\,,
\end{equation}
where $\lambda$ is a suitable affine parameter for the geodesics in question. For the global INP coordinates, we can write the geodesic equations for the Schwarzschild spacetime in terms of coordinates and the spin coefficients: \\
\newline
   \noindent
   \underline{The $\ddot{u}$ equation}:
   \begin{equation}
   \begin{aligned}
   64 M^3
   \ddot{u}+ &\rho \left(\frac{\sqrt{2} \sin ^2(\theta
   ) \left(8 M^2+u z\right)^2
   \dot{\phi}^2}{\sqrt{\frac{z}{8 M^2 u+u^2
   z}}}+\frac{\sqrt{2} \left(8 M^2+u z\right)^2
   \dot{\theta}^2}{\sqrt{\frac{z}{8 M^2 u+u^2
   z}}}\right)+\frac{256 M^5 \dot{u}
   \dot{z}}{z \left(8 M^2+u
   z\right)} \\ 
   &+ \epsilon \left(\frac{\left(256 \sqrt{2} M^4
   z^2-32 \sqrt{2} M^2 u z^3\right) (\dot{u})^2}{u
   z \left(8 M^2+u z\right) \sqrt{\frac{z}{8 M^2
   u+u^2 z}}}+ \frac{\left(-256 \sqrt{2} M^4 u z-32 \sqrt{2}
   M^2 u^2 z^2\right) \dot{z}
   \dot{u}}{u z \left(8 M^2+u z\right)
   \sqrt{\frac{z}{8 M^2 u+u^2 z}}}\right) \\ 
   &+ \frac{\left(-256
   M^5 z \sqrt{\frac{z}{8 M^2 u+u^2 z}}-64
   M^3 u z^2 \sqrt{\frac{z}{8 M^2 u+u^2 z}}\right)
   (\dot{u})^2}{u z \left(8 M^2+u z\right)
   \sqrt{\frac{z}{8 M^2 u+u^2 z}}}=0 \, .
   \end{aligned}
   \end{equation}
\newline
   \noindent
   \underline{The $\ddot{z}$ equation}:
   \begin{equation}
   \begin{aligned}	
   &64 M^3 \left(8 M^2+u z\right)
   \ddot{z}-\frac{256 M^5
   \dot{z}^2}{z} \\ 
   &+\rho \left(\frac{\sqrt{2} z
   \sin ^2(\theta ) \left(8 M^2-u z\right) \left(8 M^2+u
   z\right)^2 \dot{\phi}^2}{u \sqrt{\frac{z}{8 M^2
   u+u^2 z}}}+\frac{\sqrt{2} z \left(8 M^2-u z\right) \left(8
   M^2+u z\right)^2 \dot{\theta}^2}{u
   \sqrt{\frac{z}{8 M^2 u+u^2 z}}}\right) \\ 
   &+\epsilon
   \left(\begin{aligned}
   	\frac{\left(64 \sqrt{2} M^2 u z^5-512 \sqrt{2} M^4
   z^4\right) \dot{u}^2}{u z \left(8 M^2+u
   z\right) \sqrt{\frac{z}{8 M^2 u+u^2 z}}}+ \\ \frac{\left(96
   \sqrt{2} M^2 u z^3 \left(8 M^2+u z\right)-256 \sqrt{2}
   M^4 z^2 \left(8 M^2+u z\right)\right)
   \dot{z} \dot{u}}{u z \left(8
   M^2+u z\right) \sqrt{\frac{z}{8 M^2 u+u^2
   z}}}+\\ \frac{\left(32 \sqrt{2} M^2 u^2 z^2 \left(8 M^2+u
   z\right)+256 \sqrt{2} M^4 u z \left(8 M^2+u
   z\right)\right) \dot{z}^2}{u z \left(8 M^2+u
   z\right) \sqrt{\frac{z}{8 M^2 u+u^2 z}}}
   \end{aligned}\right) \\
   &+ \frac{\left(512 M^5 z^3 \sqrt{\frac{z}{8
   M^2 u+u^2 z}}+64 M^3 u z^4 \sqrt{\frac{z}{8
   M^2 u+u^2 z}}\right) \dot{u}^2}{u z \left(8
   M^2+u z\right) \sqrt{\frac{z}{8 M^2 u+u^2
   z}}}\\ 
   &+\frac{\left(256 M^5 z \left(8 M^2+u z\right)
   \sqrt{\frac{z}{8 M^2 u+u^2 z}}+128 M^3 u z^2
   \left(8 M^2+u z\right) \sqrt{\frac{z}{8 M^2 u+u^2
   z}}\right) \dot{u} \dot{z}}{u z
   \left(8 M^2+u z\right) \sqrt{\frac{z}{8 M^2 u+u^2
   z}}}=0  \, .
   \end{aligned}
   \end{equation}
\newline	
   \noindent
   \underline{The $\ddot{\theta}$ equation}:
   \begin{align}	\nonumber
   \ddot{\theta}-&\frac{\beta \sin^2(\theta )
   \left(8 M^2+u z\right) \dot{\phi}^2}{\sqrt{2}
   M} \\
   &+\rho \left(-\frac{z \dot{u}
   \dot{\theta}}{2 \sqrt{2} u
   \sqrt{\frac{1}{\frac{u^2}{M^2}+\frac{8
   u}{z}}}}-\frac{\dot{z} \dot{\theta}}{2
   \sqrt{2} \sqrt{\frac{1}{\frac{u^2}{M^2}+\frac{8
   u}{z}}}}\right)+\frac{u \dot{z}
   \dot{\theta}}{8 M^2+u z}+\frac{z
   \dot{u} \dot{\theta}}{8 M^2+u
   z}=0 \, .
   \end{align}
\newline
   \noindent
   \underline{The $\ddot{\phi}$ equation}:
   \begin{align}	\nonumber
   \ddot{\phi}\ +\ &\dot{\theta}
   \dot{\phi} \left(\cot (\theta )+\frac{\beta
   \left(8 M^2+u z\right)}{\sqrt{2} M}\right) \\
   &+\rho \dot{\phi} \left(-\frac{\dot{z}}{2
   \sqrt{2} \sqrt{\frac{M^2 z}{8 M^2 u+u^2 z}}}-\frac{z
   \dot{u}}{2 \sqrt{2} u \sqrt{\frac{M^2 z}{8
   M^2 u+u^2 z}}}\right)+\dot{\phi} \left(\frac{u
   \dot{z}}{8 M^2+u z}+\frac{z
   \dot{u}}{8 M^2+u
   z}\right)=0 \, .
   \end{align}
\subsection{`Standard' form of the geodesic equations}
	
Since the above is quite difficult to parse, we also list the usual form of the geodesic equations in terms of the Christoffel symbols, as in equation (\ref{geodesiceqn}):
\begin{subequations}
\begin{align}
   &\ddot{u}-\frac{u \left(8 M^2+u z\right) \left(\sin
   ^2(\theta ) \dot{\phi}^2+\dot{\theta}^2\right)}{16
   M^2}-\frac{z \left(16 M^2+u z\right)
   \dot{u}^2}{\left(8 M^2+u z\right)^2}=0, \\
   &\ddot{z}+\frac{z \left(u z-8 M^2\right) \left(\sin ^2(\theta )
   \dot{\phi}^2+\dot{\theta}^2\right)}{16
   M^2}+\frac{z^3 \left(24 M^2+u z\right)
   \dot{u}^2}{\left(8 M^2+u z\right)^3}+\frac{2 z
   \left(16 M^2+u z\right) \dot{u}
   \dot{z}}{\left(8 M^2+u
   z\right)^2}=0, \\
    &\ddot{\theta}+\frac{2 u
   \dot{z} \dot{\theta}+2 z
   \dot{u} \dot{\theta}}{8 M^2+u
   z}-\sin (\theta ) \cos (\theta ) \dot{\phi}^2=0, \\
   &\ddot{\phi}+\frac{2 u
   \dot{z} \dot{\phi}+2 z
   \dot{u} \dot{\phi}}{8 M^2+u z}+2
   \cot (\theta ) \dot{\theta} \dot{\phi}=0.
\end{align}
\end{subequations}
As we can see, any divergent behaviour of the spin coefficients at the horizon is not physically realized in the geodesic equations, due to cancellations and simplifications. The only singularities in these equations are at points corresponding to $r=0$, as in the Schwarzschild coordinates system. 
%
%
\section{Fundamentals of teleparallel geometry}

The teleparallel geometry is characterized by the frame, $e^{a}{}_{\mu}$, and a metric-compatible spin connection, $\omega^{a}{}_{b\mu}$, antisymmetric in the algebraic indices, which generates non-zero torsion but vanishing curvature \cite{aldrovandi1732013}. As mentioned earlier, tetrads satisfy orthogonality conditions and establish a relationship between the spacetime metric, \( g_{\mu\nu} \), and the tangent-space metric, \( \eta_{ab} \), as given by Eq. (\ref{gmetric}). Now, we prefer to use the diagonal metric:
\begin{equation}\label{Eta}
    \eta_{ab} = \begin{pmatrix}
        -1 & 0 & 0 & 0 \\
        0 & 1 & 0 & 0 \\
        0 & 0 & 1 & 0 \\
        0 & 0 & 0 & 1
    \end{pmatrix} \, ,
\end{equation}
instead of (\ref{eta}). Note that (\ref{eta}) and (\ref{Eta}) are related by a simple basis transformation. The total covariant derivative that operates on both tangent-space and spacetime indices vanishes when acting on the tetrad (such a condition is known as the ``{\it tetrad postulate}" \cite{bahamonde862023}) and can be used to define the teleparallel connection 
\begin{equation}\label{Gamma}
\Omega^{\rho}{}_{\nu\mu}=e_{a}{}^{\rho}\partial_{\mu}e^{a}{}_{\nu}+e_{a}{}^{\rho}\, \omega^{a}{}_{b\mu}\, e^{b}{}_{\nu}\, .
\end{equation} 
This connection exhibits the tensorial property known as torsion, defined as
\begin{equation}\label{torsion}
T^{\sigma}{}_{\mu\nu}=\Omega^{\sigma}{}_{\nu\mu}-\Omega^{\sigma}{}_{\mu\nu} \, .
\end{equation}
Such a tensor can be decomposed into irreducible parts under the global Lorentz group \cite{aldrovandi1732013}. These parts include the vectorial, axial, and purely tensorial components, denoted as
\begin{equation}\label{IPT}
\mathscr{V}_{\mu}=T^{\nu}{}_{\nu\mu} \,,\quad 
\mathscr{A}_{\mu}=\frac{1}{6}\varepsilon_{\mu\nu\rho\sigma}\,T^{\nu\rho\sigma} \, , \quad 
\mathscr{T}_{\sigma\mu\nu}=T_{\sl\sigma\mu\sr\nu}+\frac{1}{3}\left(g_{\sigma\al\nu}\mathscr{V}_{\mu\ar}+g_{\mu\al\nu}\mathscr{V}_{\sigma\ar} \right) \, ,
\end{equation} 
respectively. The axial torsion is defined using the totally antisymmetric Levi-Civita tensor, $\varepsilon_{\mu\nu\rho\sigma}$, associated with metric, $g_{\mu\nu}$. 
\subsection{Connection decomposition}
An important theorem arises from the metric-preserving property of Lorentz connections, which states that any Lorentz connection can be decomposed into the Levi-Civita connection and the contortion tensor \cite{aldrovandi1995introduction}; that is,
\begin{equation}\label{ConDec}
\Omega^{\sigma}{}_{\mu\nu}= \Gamma^{\sigma}{}_{\mu\nu}+K^{\sigma}{}_{\mu\nu} \ ,
\end{equation}
where
\begin{equation}\label{Cont}
K^{\sigma}{}_{\mu\nu}=\frac{1}{2}\left(T_{\nu}{}^{\sigma}{}_{\mu} +T_{\mu}{}^{\sigma}{}_{\nu}-T^{\sigma}{}_{\mu\nu}\right)
\end{equation}
is the contortion tensor. A consequence of the decomposition (\ref{ConDec}) is that the curvature tensor associated with the Levi-Civita connection can be expressed as
\begin{equation}\label{RimanToR}
\mathcal{R}^{\rho}{}_{\sigma\mu\nu}=N^{\rho}{}_{\sigma\mu\nu}-N^{\rho}{}_{\sigma\nu\mu} \ ,
\end{equation}
where
\begin{equation}
N^{\rho}{}_{\sigma\mu\nu}\equiv K^{\rho}{}_{\sigma\mu|\nu}+K^{\rho}{}_{\lambda\mu}K^{\lambda}{}_{\sigma\nu}+K^{\rho}{}_{\sigma\lambda}K^{\lambda}{}_{\mu\nu} \, .
\end{equation}
Calculating the Kretschmann scalar $\mathcal{K}$ on the left-hand side of equation (\ref{RimanToR}) yields:
\begin{equation}\label{Krets}
\mathcal{K}= 2(N_{\rho\sigma\mu\nu}-N_{\rho\sigma\nu\mu})N^{\rho\sigma\mu\nu}\, .
\end{equation}
We can examine the explicit terms on the right-hand side of Eq.(\ref{Krets}) and express them in terms of the contortion tensor as follows:
\begin{alignat}{1}\label{Nt1}\nonumber
N_{\rho\sigma\mu\nu}N^{\rho\sigma\mu\nu}&=K_{\rho\sigma\mu|\nu}K^{\rho\sigma\mu|\nu}+2K^{\mu\nu\rho}(K_{\rho}{}^{\sigma\alpha}K_{\mu\nu\sigma|\alpha}+K_{\nu}{}^{\sigma\alpha}K_{\mu\sigma\rho|\alpha}) \\[1.7ex]
&\quad +K^{\mu\nu\rho}(2K_{\alpha\rho\beta}K_{\mu}{}^{\sigma\alpha}K_{\nu\sigma}{}^{\beta}+K_{\sigma\alpha\beta}(K_{\mu}{}^{\sigma}{}_{\rho}K_{\nu}{}^{\alpha\beta}+K_{\mu\nu}{}^{\sigma}K_{\rho}{}^{\alpha\beta})) \, ,
\end{alignat}
and
\begin{alignat}{1}\label{Nt2}\nonumber
N_{\rho\sigma\mu\nu}N^{\rho\sigma\nu\mu}&=K_{\rho\sigma\mu|\nu}K^{\rho\sigma\nu|\mu}+2K^{\mu\nu\rho}(K_{\nu}{}^{\sigma\alpha}K_{\mu\sigma\alpha|\rho}+K_{\rho}{}^{\sigma\alpha}K_{\mu\nu\alpha|\sigma}) \\[1.7ex]
&\quad + K^{\mu\nu\rho}(K_{\mu\nu}{}^{\sigma}K_{\sigma}{}^{\alpha\beta}K_{\sigma\beta\alpha}+K_{\mu}{}^{\sigma\alpha}(2K_{\alpha\beta\rho}K_{\nu\sigma}{}^{\beta}+K_{\nu}{}^{\beta}{}_{\alpha}K_{\alpha\beta\rho})) \, .
\end{alignat}
We have derived a set of invariants from equations (\ref{Nt1}) and (\ref{Nt2}), constructed from the contortion tensor and its covariant derivatives. These scalars have been overlooked in existing literature, making their inclusion a crucial aspect of our approach.
\subsection{Teleparallel scalars}

Teleparallel scalars play a crucial role in the Lagrangian formulations of teleparallel theories. In the context of TEGR, the Lagrangian density depends on the scalar given by 
\begin{equation}\label{TdeS}
T=\frac{1}{2}S_{\rho}{}^{\mu\nu}T^{\rho}{}_{\mu\nu} \, ,
\end{equation}
where $S_{\rho}{}^{\mu\nu}$ is the superpotential, which is antisymmetric in the last two indices and takes on the form
\begin{equation}\label{S}
S_{\rho}{}^{\mu\nu}=K^{\mu\nu}_{\rho}+\delta^{\mu}_{\rho}T^{\lambda\nu}{}_{\nu}-\delta^{\nu}_{\rho}T^{\lambda\mu}{}_{\mu} \, .K_{ab}{}^{c}
\end{equation}
The invariant (\ref{TdeS}) can also be expressed in terms of the irreducible parts of the torsion (\ref{IPT}) as follows:
\begin{equation}\label{TorS2}
T=\frac{3}{2}\mathscr{A}-\frac{2}{3}\mathscr{V}+\frac{2}{3}\mathscr{T} \, ,
\end{equation}
where
\begin{equation}\label{Tds}
\mathscr{A}=\mathscr{A}^{\mu}\mathscr{A}_{\mu}\, , \quad \mathscr{V}=\mathscr{V}^{\mu}\mathscr{V}_{\mu} \, , \quad \mathscr{T}=\mathscr{T}^{\sigma\mu\nu}\mathscr{T}_{\sigma\mu\nu} \, .
\end{equation}
There are many higher-order invariants derived from expressions (\ref{Nt1}) and (\ref{Nt2}) that we can consider. We will focus on two specific ones involving covariant derivatives that can be written in terms of the torsion tensor; using (\ref{Cont}), we get
\begin{equation}\label{DK1}
K_{\rho\sigma\mu|\nu}K^{\rho\sigma\mu|\nu}=\frac{1}{2} T_{\sigma\mu\rho|\nu}T^{\mu\sigma\rho|\nu}+\frac{3}{4}T_{\mu\sigma\rho|\nu}T^{\mu\sigma\rho|\nu} \, ,
\end{equation}
and
\begin{equation}\label{DK2}
K_{\rho\sigma\mu|\nu}K^{\rho\sigma\nu|\mu}=\frac{1}{4}T_{\sigma\nu\mu|\rho}T^{\rho\nu\mu|\sigma}+\frac{1}{2}\left(T_{\mu\nu\sigma|\rho}-T_{\nu\mu\sigma|\rho} \right)T^{\mu\nu\rho|\sigma}-T_{\nu\mu\sigma|\rho}T^{\rho\nu\mu|\sigma} \, .
\end{equation}
We aim to leverage these scalar invariants to gain further insights into the presence or absence of horizons within exact solutions of vacuum FE.

To investigate static and spherically symmetric solutions in TG, we first need to display the general forms for the the frame (or co-frame) and the flat, metric compatible spin connection that respect the assumed static and spherically symmetric (affine) symmetries  \cite{coley616439,coley842024}. 
\subsection{Static spherical symmetric spacetimes}

In teleparallel gravity, stationary or static spherically symmetric geometries have been explored in various applications \cite{pfeifer2021}. Working in coordinates $x^{\mu}=(t,r,\theta,\phi)$ and using the affine frame symmetry generators of the three-dimensional spherical symmetry group \cite{coley616439}, the most general class of these geometries was derived by ``diagonalizing" the frame into a tetrad defined by three arbitrary functions: $A_{1}(t,r)$, $A_{2}(t,r)$, and $A_{3}(t,r)$. This invariant symmetry frame allows for the derivation of the most general metric-compatible connection. By imposing the flatness condition, any spherically symmetric teleparallel geometry can be described using these three functions, along with two additional functions, $\chi(t,r)$ and $\psi(t,r)$, in the spin connection components \cite{coley616439}. Here we are interested in the static case, and using the additional affine static spherically symmetric generator $c^{-1}_{\n}\partial_{t}$, the frame components are found to be
\begin{equation}\label{TetradSS}
e^{a}{}_{\mu}=
\begin{pmatrix}
A_1(r) & 0 & 0 &0\\
0 & A_2(r) & 0 & 0 \\
0 & 0 & A_3(r) & 0 \\
0 & 0 & 0 & A_3(r)\,\sin \theta
\end{pmatrix} ,
\end{equation}
while the non-zero spin connection components are given by
\begin{alignat}{1}\label{SpinV}\nonumber
&\omega_{133}=\omega_{144}=\frac{\cos\chi(r)\sinh\psi(r)}{A_{3}(r)} \, , \quad  \omega_{134}=\omega_{413}=\frac{\sin\chi(r)\sinh\psi(r)}{A_{3}(r)} \, ,\\[1.5ex]
& \omega_{234}=\omega_{423}=\frac{\sin\chi(r)\cosh\psi(r)}{A_{3}(r)} \, , \quad  \omega_{233}=\omega_{244} =\frac{\cos\chi(r)\cosh\psi(r)}{A_{3}(r)}\,,\\[1.5ex]\nonumber
&\omega_{212}=\frac{\partial_{r}\psi(r)}{A_{2}(r)} \, , \quad \omega_{432}=\frac{\partial_{r}\chi(r)}{A_{2}(r)} \, , \quad \omega_{434}=\frac{\cot\theta}{A_{3}(r)} \, .
\end{alignat}
It is important to note that we still have the coordinate freedom to choose 
$A_{3}(r)=r$, whenever $A_{3}(r)$ is not constant. Additionally, any choice of the arbitrary functions $\chi(r)$ and $\psi(r)$ uniquely determines a teleparallel geometry, as any alteration in the spin connection that affects the form of $\chi(r)$ or $\psi(r)$ also changes the tetrad.
%
%
\section{TEGR}

TEGR is dynamically equivalent to GR, in that their FE and classical predictions are identical \cite{aldrovandi1732013}. In GR, the Einstein-Hilbert Lagrangian density depends on the metric tensor, which describes curvature through the Riemann tensor associated with the Levi-Civita connection. In contrast, the TEGR Lagrangian density depends on the tetrad and the teleparallel spin connection, which is metric-compatible, flat, and exhibits torsion (the quantity that describes the gravitational interactions in that theory \cite{bahamonde862023}). The Lagrangians of both theories differ only by a boundary term, ensuring dynamical equivalence between their actions. The TEGR action is given by
\begin{equation}
\mathcal{S}= -\frac{1}{2\kappa} \int e \left( T-2\kappa \mathcal{L}_{m} \right) d^{4}x \, ,
\end{equation}
where $e$ is the tetrad determinant, $\kappa=8\pi G/c^4$ is the gravitational coupling constant, with units chosen so that $c=1$, and $\mathcal{L}_{m}$ is the matter Lagrangian density. Variation with respect to the tetrad yields the FE
\begin{equation}\label{FETEGR}
W_{\mu\nu}\equiv\frac{1}{2}g_{\mu\nu}T+e^{a}{}_{\mu}\left(S_{a\nu}{}^{\rho}_{~|\rho}+\omega^{b}{}_{a\rho}S_{b}{}^{\rho}{}_{\nu} \right)+\frac{1}{2}T_{\nu\sigma}{}^{\rho}S_{\mu}{}^{\sigma}{}_{\rho}+T^{\rho}{}_{\sigma\rho}S_{\mu\nu}{}^{\sigma}-T^{\sigma}{}_{\mu\rho}S_{\sigma\nu}{}^{\rho} =\kappa\Theta_{\mu\nu} \, ,
\end{equation}
where the energy-momentum tensor is given by
\begin{equation}\label{momento}
\Theta_{\mu\nu}=-\frac{2}{e}\frac{\delta(e\mathcal{L}_{m})}{\delta e^{a}{}_{\rho}}e^{a}{}_{\mu} \, g_{\rho\nu} \, .
\end{equation}
Assuming minimal, Lorentz-invariant matter coupling makes the energy-momentum tensor $\Theta_{\mu\nu}$ symmetric \cite{bahamonde862023}, allowing the FE (\ref{FE}) to be separated into symmetric and antisymmetric parts
\begin{equation}\label{FESA}
W_{\sl\mu\nu\sr}  = \kappa\Theta_{\sl\mu\nu\sr} \ , \quad \text{and}\quad W_{\al\mu\nu\ar} = 0 \,,
\end{equation}
where the antisymmetric equations $W_{\al\mu\nu\ar}$ are identically satisfied in TEGR.
\subsection{TEGR Schwarzschild-like solution}
In the static spherically symmetric case, the symmetric FE of TEGR do not receive any contributions from the spin connection components, while the antisymmetric FE are identically satisfied. Consequently, there are no constraints on the functions \(\chi(r)\) and \(\psi(r)\), which represent Lorentz degrees of freedom that are not apparent in GR.
By choosing $A_{3}(r)=r$, and assuming $\chi(r)=(2n+1)\pi$ (with $n\in\mathbb{Z}^+ $) and $\psi(r)=0$ without loss of generality within the ansatz provided by (\ref{TetradSS}) and (\ref{SpinV}), the FE (\ref{FETEGR}) in vacuum yield the Schwarzschild tetrad  
\begin{equation}\label{hs}
e^{a}{}_{\mu}=
\begin{pmatrix}
\sqrt{1-2M/r} & 0 & 0 &0\\
0 & 1/\sqrt{1-2M/r} & 0 & 0 \\
0 & 0 & r & 0 \\
0 & 0 & 0 & r\sin\theta
\end{pmatrix} .
\end{equation}
Here, we prefer to work in an orthonormal frame, which differs from the frame used earlier (Eq. \ref{frameGR}) by a basis transformation. Equation (\ref{hs}) leads to the GR Schwarzschild metric (\ref{schm}) via (\ref{gmetric}). The corresponding non-zero components of the spin connection are:
\begin{equation}\label{os}
\omega_{323}=\omega_{424}=1/r , \quad \omega_{434}=\cot\theta/r \, .
\end{equation}
The tetrads are defined up to local Lorentz transformations, which are transformations of the tangent-space coordinates \cite{aldrovandi1732013}. Under such a transformation, the tetrad and the spin connection transform as follows
\begin{equation}\label{LTr}
e^{\prime a}{}_{\mu}=\Lambda_{b}{}^{a}e^{b}{}_{\mu}  \, , \quad \omega^{\prime a}{}_{b\mu}=\Lambda_{c}{}^{a}\omega^{c}{}_{e\mu}\Lambda^{e}{}_{b} +\Lambda_{c}{}^{a}\partial_{\mu}\Lambda^{c}{}_{b} \, ,
\end{equation}
where the Lorentz transformation satisfies $\Lambda^{a}{}_{c}\Lambda_{b}{}^{c}=\delta^{a}_{b}$. Choosing  the Lorentz transformation to be 
\begin{equation}\label{LT}
\Lambda^{a}{}_{b}=
\begin{pmatrix}
1 & 0 & 0 & 0 \\
0 &  \sin\theta\cos\phi & \sin\theta\sin\phi & \cos\theta \\
0 & \cos\theta\cos\phi & \sin\theta\sin\phi & -\sin\theta \\
0 & -\sin\phi & \cos\phi & 0
\end{pmatrix} ,
\end{equation}
allows us to find the tetrad in the proper frame given by
\begin{equation}\label{hps}
e^{\prime a}{}_{\mu}= 
\begin{pmatrix}
\sqrt{1-2M/r} & 0 & 0 &0\\
0 & \sin\theta\cos\phi/\sqrt{1-2M/r} & r\cos\theta\cos\phi & -r\sin\theta\sin\phi \\
0 & \sin\theta\sin\phi/\sqrt{1-2M/r}& r\cos\theta\sin\phi & r\sin\theta\cos\phi \\
0 & \cos\theta/\sqrt{1-2M/r} & -r\sin\theta & 0
\end{pmatrix} ,
\end{equation}
in which the spin connection components are zero, $ \omega^{\prime a}{}_{b\mu} = 0$. This tetrad and spin connection pair also satisfies the TEGR FE (\ref{FETEGR}) in vacuum.

Note that (\ref{hs}) and (\ref{hps}) are valid solutions for $r>2M$. We are interested in analyzing the behavior of these static and spherically symmetric solutions near $r=2M$. For this, we will determine the value of some of the torsion invariants. Due to its presence in the TEGR Lagrangian, $T$ is the most representative scalar and is given by
\begin{equation}\label{TsV}
T=-\frac{4\left(M-r+\sqrt{r(r-2M)}\right)}{r^{2}\sqrt{r(r-2M)}} \, .
\end{equation}
It is also useful to consider the values of the scalars given by the irreducible components of torsion (\ref{Tds}), which are given by
\begin{equation}\label{TdecV}
\mathscr{A}=0 \, , \quad \mathscr{V}=\frac{\left(3M-2r+2\sqrt{r(r-2M)}\right)^2}{r^{3}(r-2M)} \, , \quad \mathscr{T}=\frac{\left(3m-r+\sqrt{r(r-2M)}\right)^2}{r^{3}(r-2M)} \, .
\end{equation}
Additionally, we include the supplementary scalars given by Eq.(\ref{DK1}), which was found to be
\begin{alignat}{1}\label{DK1V}\nonumber
K_{\rho\sigma\mu|\nu}K^{\rho\sigma\mu|\nu} &=\frac{1}{r^{6}(r-2M)^{2}}\left(24(1-\sqrt{1-2M/r})r^4 +24m(5\sqrt{1-2M/r}-6)r^3 \right. \\[1.5ex]
&\quad \left.  +48M^{2}(7-4\sqrt{1-2M/r})r^2+16M^3(6\sqrt{1-2M/r}-23)r+ 162M^4  \right) \, ,
\end{alignat}
and Eq.(\ref{DK2}), given by
\begin{equation}\label{DK2V}
K_{\rho\sigma\mu|\nu}K^{\rho\sigma\nu|\mu}=-\frac{24}{r^6}\sqrt{r(r-2M)}\left(M-r\sqrt{r(r-2M)} \right) \, .
\end{equation}
As \( r \) approaches \( 2M \), all of the invariants listed above diverge, except for \(\mathscr{A}\), which vanishes, and Eq. (\ref{DK2V}), which approaches zero in this limit. Therefore, the invariants exhibit singularities at (both \( r = 0 \) and) \( r = 2M \).

For comparison, let us reconsider the GR perspective. Using Eq.~(\ref{gmetric}) and either Eq.~(\ref{hs}) or Eq.~(\ref{hps}), we obtain the Schwarzschild metric as given by Eq.~(\ref{schm}). The key invariants associated with the Levi-Civita connection in GR are the Ricci scalar, which vanishes for all vacuum solutions, and the Kretschmann scalar given by Eq.~(\ref{Ksch}), which indicates that there is no singularity at \(r=2M\). Additionally, we have the scalar
\begin{equation}
\mathcal{C}=\mathcal{R}_{\rho\sigma\mu\nu;\alpha}\mathcal{R}^{\rho\sigma\mu\nu;\alpha}= 720 \frac{M^2}{r^8}\left(1-\frac{2M}{r}\right) \, ,
\end{equation}
which identifies the horizon at \(r=2M\) via its root. Note that, from this perspective, there is only a singularity at \(r=0\) and a horizon at \(r=2M\). In contrast, the teleparallel invariants also exhibit a singular surface $r=2M$.

\subsection{TEGR Kerr-like solution}

A static and axially symmetric tetrad that satisfies the TEGR vacuum FE can be found through a coordinate transformation \cite{pereira2001axial} and is given by
\begin{equation}\label{hk}
e^{a}{}_{\mu}=
\begin{pmatrix}
\Sigma/\rho & 0 & 0 &2 M l r\sin^{2}\theta/\rho\Sigma\\
0 & \rho/\Delta & 0 & 0 \\
0 & 0 & \rho & 0 \\
0 & 0 & 0 & \Delta\rho \sin\theta/\Sigma
\end{pmatrix} .
\end{equation}
The corresponding spin connection is found to be
\begin{equation}\label{ok}
\omega_{323}=1/\rho , \quad \omega_{424}=\Sigma/\rho\Delta , \quad \omega_{434}=\Sigma\cot\theta/\rho\Delta \, ,
\end{equation}
where
\begin{equation}\label{fun}
\Delta=\sqrt{r^{2}+l^{2}-2Mr} \, , \quad \rho=\sqrt{r^{2}+l^{2}\cos^{2}\theta} \, , \quad \Sigma=\sqrt{\rho^{2}-2Mr} \, .
\end{equation}
Using the transformation rules (\ref{LTr}) and the Lorentz transformation given in (\ref{LT}), it is possible to find the corresponding tetrad in the proper frame given by
\begin{equation}\label{hpk}
e^{\prime a}{}_{\mu}= 
\begin{pmatrix}
\Sigma/\rho & 0 & 0 &-2Mlr\sin^{2}\theta/\rho\Sigma\\
0 & \rho\sin\theta\cos\phi/\Delta & \rho\cos\theta\cos\phi & -\rho\Delta\sin\theta\sin\phi/\Sigma \\
0 & \rho\sin\theta\sin\phi/\Delta & \rho\cos\theta\sin\phi & \rho\Delta\sin\theta\cos\phi/\Sigma \\
0 & \rho\cos\theta/\Delta & -\rho\sin\theta & 0
\end{pmatrix} ,
\end{equation}
We verified that both (\ref{hk}-\ref{ok}) and (\ref{hpk}) satisfy the TEGR FE (\ref{FE}) in vacuum. Note that these solutions are valid for $\rho>\sqrt{2Mr}$. Now we are interested in analyzing the behavior of these solutions near the roots of  $\Delta=0$, which are given by
\begin{equation}\label{rpm}
r_{\pm}=M\pm\sqrt{M^{2}-l^{2}} \, ,
\end{equation}
and represent the locations of the inner and outer horizons in GR. Evaluating the functions in (\ref{fun}) at $r=r_{\pm}$, we obtain
\begin{equation}
\Delta_{\pm}=0 \, , \quad \rho_{\pm}=\sqrt{r^2_{\pm}+l^2\cos^2\theta} \, , \quad \Sigma_{\pm}=\sqrt{\rho_{\pm}^2-2Mr_{\pm}} .
\end{equation}
To gain insights into the features of the teleparallel geometry at $r = r_{\pm}$, we compute several teleparallel invariants. Beginning with the torsion scalar $T$ defined in (\ref{TorS2}), we find
\begin{alignat}{1}\nonumber\label{TsVK}
T&=\frac{1}{\Delta\rho^{6}\Sigma^{4}} \left(-2\rho^{4}\Sigma^{3}\left((l\cos\theta)^{2}+\Sigma(M-r+\Sigma) \right)+\Delta\left(2(l\sin\theta)^{2}\left(2Mr(r-M) \right.\right.\right. \\[0.5ex]\nonumber
&\quad + \left. (l\cos\theta-M)\Sigma^{2}\right) \left. \left(2Mr(M-r)+(l\cos\theta+M)\Sigma^{2}\right) + \rho^{4}\left(2\Sigma^{2}\left((l\cos\theta)^{2} \right.\right.\right. \\[0.5ex] \nonumber
&\quad -\left.(M-r)(M-r+\Sigma)\right)+ \left. \left.\left. (l^{2}\sin(2\theta))^{2}/2\right)-(l^{2}\rho\Sigma\sin(2\theta))^{2}\right) \right. \\[0.5ex]
&\quad \left. +2\Delta^{3}\left(((M-r)\rho^{2}+r\Sigma^{2})^{2}-(2lMr\cos\theta)^{2} \right) \right) \, .
\end{alignat}
It is also useful to consider the values of the scalars given by the irreducible components of torsion, which are given by
\begin{subequations}
\begin{alignat}{1}\nonumber
\mathscr{V} &=\frac{1}{\Delta^2\rho^6} \left(\rho^{4}((2\Delta+M)(M+2\Sigma)+(\Delta-\Sigma)^{2}\csc^{2}\theta) +r(\Delta^{2}+\rho^{2})\left(r(\Delta^{2}+\rho^{2}) \right. \right.\\[0.5ex]
& \quad\left.\left. -2\rho^{2}(\Delta+M+\Sigma) \right)+\Delta l^{2}\cos^{2}\theta(2\rho^{2}(\Sigma-\Delta)+\Delta l^{2}\sin^{2}\theta) \right) \, ,\\[1.7ex]
\mathscr{A}&=\frac{4M^2l^2}{9\rho^{6}\Sigma^4}((2(M-r)r+\Sigma^2)^{2}\sin^{2}\theta+(2r\Delta\cos\theta)^{2}) \, , \\[1.7ex]\nonumber
\mathscr{T}&=\frac{1}{\Delta^2\rho^{6}\Sigma^{4}} \left(\Sigma^{4}(\rho^{4}((\Sigma\csc\theta)^{2}+(M-r)(M-r+2\Sigma))-(4lMr\cos\theta)^{2}) \right.\\[0.5ex]\nonumber
& \quad-(2l^{2}Mr\sin(2\theta))^{2} \left(\Sigma^{2}+\Delta^{2}\right) -\Delta\rho^{2}\Sigma^{3}\left(\rho^{2}\left(3(l\cos\theta)^{2}+\Sigma\left(M-r\right.\right. \right. \\[0.5ex]\nonumber
&\quad \left.\left.\left.+(1+2\cot^{2}\theta)\Sigma\right)\right)-2(l\cos\theta\Sigma)^{2} \right) +\Delta^2\left(\rho^{2}\Sigma^{4}\left((\rho\csc\theta)^{2}-2r(M-r+\Sigma)\right) \right. \\[0.5ex] \nonumber
& \quad +3\rho^{4}\Sigma^{2}\left((l\cos\theta)^{2}-(M-r)(M-r+\Sigma)\right) + \left.(l\sin\theta)^{2}\left(3(l\rho\cos\theta)^{2}(\rho^{2}-2\Sigma^{2}) \right. \right. \\[0.5ex]\nonumber
& \quad +2\Sigma^{4}(2(l\cos\theta)^{2}-2M^{2}-(\rho\cot\theta)^{2})- \left. \left. 16M^{2}r(M-r)((M-r)r+\Sigma^{2}) \right) \right) \\[0.5ex]
&\quad \left. -2r\Delta^{3}\rho^{2}\Sigma^{4}+\Delta^{4}\left(3(M-r)\rho^{2}\left((M-r)\rho^{2} +  2r\Sigma^{2}\right)+4r^{2}\Sigma^{4} \right) \right)\, .
\end{alignat}
\end{subequations}
The forms of the supplementary invariants (\ref{DK1}) and (\ref{DK2}) are too complicated to be included explicitly here; instead, we are going to write down their evaluation at $r=r_{\pm}$, which were found to be
\begin{alignat}{1}\nonumber
K_{\rho\sigma\mu|\nu}K^{\rho\sigma\mu|\nu}|_{r_{\pm}} &= \frac{(l^{2}-4Mr_{\pm}-l^{2}\cos(2\theta))^{8}}{256\Delta_{\pm}^{4}\rho_{\pm}^{20}}\left(4l^{4}+2M^{4}-l^{2}(9M^{2}-10Mr_{\pm}+5r_{\pm}^{2}) \right. \\[1.5ex]
&\quad + \left. (M-r_{\pm})(l^{2}(r_{\pm}-M)\cos(2\theta)+4(M^{2}-2l^{2})\Sigma_{\pm}) \right) \rightarrow \infty \, ,
\end{alignat}
and 
\begin{alignat}{1}\nonumber
K_{\rho\sigma\mu|\nu}K^{\rho\sigma\nu|\mu}|_{r_{\pm}}&=\frac{(l^{2}\cos(2\theta)+4Mr_{\pm}-l^{2})^{5}}{32\Delta_{\pm}^{4}\rho_{\pm}^{14}}\left(l^{2}(3M^{2}-2Mr_{\pm}+r_{\pm}^{2}) -4l^{4}\sin^{2}\theta\right.\\[1.7ex]
&\quad \left.-l^{2}(5M^{2}-6Mr_{\pm}+3r_{\pm}^{2})\cos(2\theta)+4(2l^{2}-M^{2})(M-r_{\pm})\Sigma_{\pm} \right) \rightarrow \infty \, .
\end{alignat}
Similarly, we can analyze the behaviour of Eq.(\ref{TsVK}) as r approaches $r_{\pm}$ which yields
\begin{equation}
T |_{r_{\pm}}= \frac{2\csc^{2}\theta}{\Delta_{\pm}\rho^{2}_{\pm}}\left((r_{\pm}-M)\sin^{2}\theta+\Sigma_{\pm}\cos(2\theta) \right) \rightarrow \infty \, ,
\end{equation}
and also for the irreducible parts of the torsion as follows
\begin{subequations}
\begin{alignat}{1} 
\mathscr{V}|_{r_{\pm}} &=  \frac{1}{\Delta_{\pm}^{2}\rho_{\pm}^{2}}\left((M-r_{\pm})(M-r_{\pm}+2\Sigma_{\pm})-l^{2} \right) \rightarrow \infty  \, ,\\[1.7ex]
\mathscr{A}|_{r_{\pm}}&= -\frac{4M^2}{9\rho_{\pm}^{6}\Sigma_{\pm}^2}(2(M-r_{\pm})r_{\pm}+\Sigma_{\pm}^2)^{2} \, , \\[1.7ex] 
\mathscr{T}|_{r_{\pm}}&= \frac{1}{\Delta_{\pm}^{2}\rho_{\pm}^{2}}
\left((M-r_{\pm})(M-r_{\pm}+2\Sigma_{\pm})-l^{2} \right) \rightarrow \infty \, .
\end{alignat}
\end{subequations}
Clearly, all teleparallel invariants diverge at $r = r_{\pm}$ except for $\mathscr{A}$. Moreover, we observed that all teleparallel invariants except $\mathscr{V}$ also diverge at the roots of $\Sigma = 0$, which corresponds to the localization of the ergosurface in GR.
For comparison, consider the GR point of view. Using Eq.(\ref{gmetric}) and either (\ref{hs}) or (\ref{hps}), we obtain the Kerr metric
\begin{equation}
g_{\mu\nu}=
\begin{pmatrix}
-(\Sigma/\rho)^{2} & 0 & 0 &2Mlr(\sin\theta/\rho)^{2}\\
0 & (\rho/\Delta)^{2} & 0 & 0 \\
0 & 0 & \rho^{2} & 0 \\
2Mlr(\sin\theta/\rho)^{2} & 0 & 0 & L^{2}\sin^{2}\theta
\end{pmatrix} ,
\end{equation}
where
\begin{equation}
L^{2}=l^{2}+r^{2} + 2l^{2}Mr\sin^{2}\theta /\rho^{2} .
\end{equation}
The representative invariants associated with the Levi-Civita connection in GR are the Ricci scalar,
\begin{equation}
\mathcal{R}=0 \, ,
\end{equation}
which vanishes, and the Kretschmann scalar
\begin{equation}
\mathcal{K}=\frac{48M^2}{\rho^{12}}\left(r^6-15r^4l^2\cos^2\theta+15r^2l^4\cos^4\theta-l^6\cos^6\theta \right) 
\end{equation}
and the scalar
\begin{equation}
\mathcal{C}=720\frac{M^2}{\rho^{18}} \Sigma^2(r^{8}-28 r^{6}l^{2}\cos^{2}\theta +70r^{4}l^{4}\cos^{4}\theta-28r^{2}l^{6}\cos^{6}\theta+l^{8}\cos^{8}\theta) 
\end{equation}
which do not diverge at $r=r_{\pm}$ or $\Sigma=0$.
%
%
\section{$F(T)$ teleparallel gravity}

The complete Lagrangian for $F(T)$ teleparallel gravity is
\begin{equation}\label{lagrangian}
\mathcal{L}= \frac{e}{2\kappa}F(T)+\mathcal{L}_{m} \, ,
\end{equation}
where $F(T)$ is a function of the scalar torsion, $T$. Variations of the Lagrangian, which include a non-trivial spin-connection \cite{krssak362019}, yield {\it Lorentz covariant} FEs, and we can use  Lagrange multipliers to impose the conditions that the spin connection is both metric compatible and flat.

The invariance of the FEs under Lorentz transformations implies that the canonical energy momentum is symmetric and it can be shown that the usual metrical energy momentum $T_{ab}$ is related to the symmetric part of the canonical energy momentum.
Variations of the Lagrangian (\ref{lagrangian}) with respect to the co-frame yield FEs that can be decomposed into symmetric and antisymmetric parts:
\begin{subequations}\label{FE}
\begin{align}
\kappa \Theta_{(ab)}&= F''(T)S_{(ab)}^{\phantom{(ab)}\nu} \partial_{\nu} T+F'(T)\mathcal{G}_{ab}  + \frac{1}{2}g_{ab}\left(F(T)-TF'(T)\right),\label{SYM_FE}\\
             0      &= F''(T)S_{[ab]}^{\phantom{[ab]}\nu} \partial_{\nu} T,\label{ASYM_FE}
\end{align}
\end{subequations}
where $\mathcal{G}_{ab}$ is the Einstein tensor computed from the Levi-Civita connection of the metric.

Note that if $T=T_0$, a constant, then the FEs for $F(T)$ teleparallel gravity are equivalent to a rescaled version of TEGR \cite{krssak362019}. In the case of TEGR, where $F(T)=T$, eqn. \eqref{ASYM_FE} is identically satisfied, and again the theory reduces to a theory that is dynamically equivalent to GR.  In general, if $F(T)$ is non-linear and $T$ is not a constant, then the anti-symmetric part of the FEs \eqref{ASYM_FE} impose constraints on the geometry \cite{vandenHoogen:2023pjs}.
\subsection{Vacuum spacetimes}

Let us consider the governing eqns. in vacuum \footnote{All of the results below are taken directly from \cite{coley616439,coley842024}, which subsume all previous results which are reviewed in \cite{bahamonde862023,cai792016} and the references therein.}. The antisymmetric FEs 
can be solved to obtain
$\sin \chi = 0$, $\cos \chi = \delta =\pm 1$ and $\psi=0$.  The symmetric eqns. can be then written:
\begin{subequations}
\begin{eqnarray}
F''\left(T\right)\,\left(\partial_r\,T\right)\,k_1\left(r\right)&=& F'\left(T\right)\,g_1\left(r\right), \label{2010a}
\\
F''\left(T\right)\,\left(\partial_r\,T\right)\,k_2\left(r\right)&=& F'\left(T\right)\,g_2\left(r\right),  \label{2010b}
\\
\frac{F\left(T\right)}{4} &=& -F''\left(T\right)\,\left(\partial_r\,T\right)\,k_3\left(r\right)+F'\left(T\right)\,g_3\left(r\right) . \label{2010c}
\end{eqnarray}
\end{subequations}
where the functions $g_1, g_2, g_3, k_1, k_2, k_3$ are explicitly given in \cite{coley842024}. From the eqns. \eqref{2010a} - \eqref{2010c} we obtain the relation $g_1\,k_2=g_2\,k_1$ and the expression
\begin{align}\label{2013}
& \frac{{k_1\left(r\right)}}{4\left[{k_1\left(r\right)} g_3\left(r\right)- {g_1\left(r\right)\,k_3\left(r\right)}\right]} = \frac{d}{dT}\left(\ln F(T)\right) .
\end{align}
\subsubsection{Specific coordinate choice}

If \( A_3 \) is not a constant\footnote{If \( A_3 = \text{const.} \), we immediately obtain \( g_2 = 0 \) and \( k_2 = \delta/A_2 \neq 0 \). From the FE \eqref{2010b}, this implies  \( F''(T)\,(\partial_r T) = 0 \), so \( F(T) \) is either linear in \( T \) (as in GR) or \( T = \text{constant} \).},  we can redefine the \( r \) coordinate to set \( A_3(r) = r \). The component functions then become:
\begin{subequations}
\begin{align}
g_1 =& \frac{1}{A_1\,A_2^3\,r^2}\Bigg[-A_2\,r^2\,\left(\partial_r^2\,A_1\right)+A_1\,A_2+\partial_r\,\left(A_1\,A_2\right)\,r+r^2\left(\partial_r\,A_1\right)\left(\partial_r\,A_2\right)-A_1\,A_2^3\Bigg] , \label{2014a}
\\
g_2 =& \frac{1}{A_1\,A_2^3\,r}\Bigg[\partial_r\,\left(A_1\,A_2\right)\Bigg] ,\label{2014b}
\\
g_3 =& \frac{1}{A_1\,A_2^3\,r^2}\Bigg[-A_1\,A_2-A_2\,r\left(\partial_r\,A_1\right)-\delta\,A_1\,A_2^2+A_1\,r\,\left(\partial_r\,A_2\right)-\delta\,A_2^2\,r\left(\partial_r\,A_1\right)\Bigg], \label{2014c}
\\
k_1 =& \frac{1}{A_1\,A_2^3\,r^2}\left[A_1\,A_2\,r+A_2\,r^2\left(\partial_r\,A_1\right)+\delta\,A_1\,A_2^2\,r\right], \label{2014d}
\\
k_2=& k_3 =  \frac{1}{A_2^2\,r}\left[1+\delta\,A_2\right]  .  \label{2014e}
\end{align}
\end{subequations}
The torsion scalar can then be expressed as:
\begin{equation}
T(r)=-2\left(\frac{\delta}{r}+\frac{1}{A_2\,r}\right)\left(\frac{\delta}{r}+\frac{1}{A_2\,r}+2\frac{\partial_r A_1}{A_2\,A_1}\right). \label{2015}
\end{equation}  
Note that the condition
$g_1\,k_2=g_2\,k_1$ now yields
\begin{equation}
\Bigg[-r^2 \frac{\left(\partial_r^2\,A_1\right)}{A_1} +1 -{A_2}^2\Bigg]    =           
 \Bigg(\frac{1}{1 + \delta A_2}\Bigg) r^2 \frac{\left(\partial_r\,A_1\right)}{A_1}          \Bigg[ \frac{\partial_r\,A_1}{A_1}  -\delta \partial_r\,A_2 \Bigg].\label{kgkg}
\end{equation}

\subsubsection{Schwarzschild-like solutions}

The coframe in which  $A_2\left(r\right)=A^{-1}_1(r)$ and $A_3\left(r\right)=r$, corresponds to {\em{Schwarzschild}}-like solutions.
 The solutions include the Schwarzschild spacetime $A_1(r)=\sqrt{1-2M/r}$ as an explicit case. We note that in vacuum we have that $F''=0$ or $T=$ constant, which leads to the GR case (with a cosmological constant).

\subsubsection{Apparent horizon}

The condition for the existence of an AH at \( r = r_h \) is that the spin coefficient satisfies \( \rho(r_h) = 0 \) \cite{GH}. For the static spherically symmetric case under consideration we have that
\begin{equation}\label{AH1}
\rho = -\frac{\partial_r A_3}{\sqrt{2} A_2 A_3} \, , 
\end{equation}
which in the specified coordinate system, and defining $a_2 = 1/A_2$, becomes
\begin{equation}\label{AH2}
\rho = -\frac{a_2(r)}{\sqrt{2}r}, 
\end{equation}
so that the AH is characterized by $a_2(r_h)=0$. Note that in the Schwarzschild solution $a_2 = \sqrt{1 - 2M/r}$, so that $r_h = 2M$.
\subsection{Analysis}

Let us assume the existence of an AH at $r = r_h$ and
write $r = r_h + \epsilon$ where $\epsilon > 0$ tends to zero.
We assume that \footnote{In the relations presented by (\ref{eq:91}), where $p>0$ so that $a_2\to 0$ as $\epsilon \to 0$ by definition, it is implicitly assumed that the spacetime is regular outside of the horizon. If this is not that case, as $\epsilon\to 0$, $T$ begins to grow without bound until a point is reached (outside the horizon) in which the spacetime is no longer regular, which is even more problematic.}
\begin{equation}\label{eq:91}
a_2 = \epsilon ^p (\alpha_1 + \alpha_2 \epsilon), ~~A_1 = \epsilon ^q (\beta_1 + \beta_2 \epsilon),
\end{equation}
where $p>0$ and $\alpha_1 > 0$, and
$q>0$ and $\beta_1 > 0$.
To highest order in $\epsilon$, from Eq. \eqref{kgkg} we obtain
\begin{equation}
r_h ^2 q(q - (1-p)) \epsilon ^{-2} + \frac{1}{\alpha_1 ^2} \epsilon ^{-2p}=0.
\end{equation}
We note that this cannot be solved to leading order for $p>1$ ($\frac{1}{\alpha_1 ^2} \neq 0$) nor for $p=1$ ($ q^2 r_h ^2 +  \frac{1}{\alpha_1 ^2} \neq 0$), so we conclude that $p<1$ (and $\epsilon ^{-2p}$ can be neglected relative to the $o(\epsilon ^{-2})$ terms). Hence we have that
\begin{equation}
q(q - (1-p))= 0.
\end{equation}
If  $q = 0$, we obtain $p = \frac{1}{2}$, but Eq. \eqref{kgkg} then leads to inconsistences at $o(\epsilon ^{-1})$ and $o(\epsilon ^{-\frac{1}{2}})$. Hence $q > 0$, and we conclude that
\begin{equation}
0<q = (1-p) < 1.
\end{equation}
Note that from \eqref{2015}, to leading order
\begin{equation}
T(r) \sim \alpha_1 q \epsilon ^{p-1},
\end{equation}
and hence diverges as $\epsilon \rightarrow 0$ (i.e., as the AH is approached). We then evaluate Eqn. \eqref{kgkg} to next leading order. We immediately obtain $p = \frac{1}{2}$ (and hence $q = \frac{1}{2}$) and $\alpha_1 = \delta >0 $, and to $o(\epsilon ^{-1})$ we obtain the solution:
\begin{equation} \frac{3\beta_2}{\beta_1}  +  \frac{\alpha_2}{\alpha_1} +  \frac{2}{ \alpha_1 ^2 r_h^2} = 0,
\end{equation}
which serves to determine $\beta_2$. We can continue to solve Eqn. \eqref{kgkg} term by term.

Note that $a_2 = \frac{1}{A_2} \sim \epsilon ^{\frac{1}{2}} \sim A_1$, which is similar to that in the Schwarschild metric. Indeed,we also note that $\frac{g_2}{k_2} \sim \epsilon ^{\frac{1}{2}} \rightarrow 0$, so that from Eqn. \eqref{2010b} as $r \rightarrow r_h$, $F(T) \rightarrow T, F^{\prime}(T) \rightarrow const., F^{\prime\prime}(T) \rightarrow 0$ and the solutions tend to the GR  Schwarschild solution.

We conclude that in all $F(T)$ theories, {\em {if}} a AH $r=r_h$ does exist, the geometry is necessarily singular there.  Hence, no black holes can exist in vacuum in the static spherically symmetric case.

%
%
\section{Discussion}	

We have presented a complete review of the Schwarzschild vacuum spacetime, with particular emphasis on the role of scalar polynomial invariants (including differential invariants) and the  null frame approach (and the related Cartan invariants), that justifies the interpretation of the well defined Schwarzschild geometry as a black hole spacetime admitting a horizon shielding the central singular point.

We then turned to teleparallel spacetimes in which the torsion characterizes the geometry, and the scalar invariants of interest are those constructed from the torsion and its (covariant) derivatives. We studied static spherically symmetric vacuum spacetimes. We first considered the Schwarzschild-like solution in TEGR and found that, unlike the situation for their GR counterpart, the torsion scalar invariants (and, in particular, the scalar $T$) diverge at the so-called Schwarzschild horizon. Therefore, within TEGR, the horizon and its interior are not part of the manifold itself. {\footnote{This occurs when any scalar polynomial invariant constructed from the torsion and its (covariant) derivatives diverges at the horizon. Indeed, the Lagrangian in TEGR is only defined over a manifold at points for which the torsion scalar does not diverge.}} In this sense the resulting spacetime is {\em not} a black hole spacetime. This is similar to what happens in Brans-Dicke theory \cite{nick} in which the `horizon' becomes a naked singularity. We then considered TEGR Kerr-like solution and obtained similar results.

Finally, we investigated static spherically symmetric vacuum spacetimes within $F(T)$ teleparallel gravity, a generalization of TEGR. We showed that if a such a geometry exists with a `horizon', then the Torsion scalar $T$ necessarily diverges there,  and the spacetime consequently does not represent a black hole in the sense discussed above.

In future work we intend to undertake a comprehensive study of possible black holes in static spherical symmetric vacuum spacetimes in NGR \cite{diego} (an alternative generalization of TEGR), and in other torsion theories of gravity. We also intend to look at non-vacuum static spherically symmetric spacetimes in $F(T)$ gravity.
\vskip 0.5cm

%
%
{\bf{Acknowledgement}}:
AAC is supported by NSERC.

\newpage
%
%

\end{document}